\documentclass[12pt]{elsarticle}
\usepackage[OT4]{fontenc}
\usepackage{hyperref} 
\usepackage{graphicx} 

\begin{document}
\begin{frontmatter}
\title{Inferring cultural regions from correlation networks of given baby names}

\author[agh]{Mateusz Pomorski}
\author[agh]{Ma{\l}gorzata J. Krawczyk}
\author[agh]{Krzysztof Ku{\l}akowski}
\author[pan]{Jaros{\l}aw Kwapie\'n}
\author[aus1,aus2,aus3]{Marcel Ausloos}

\address[agh]{AGH University of Science and Technology, Faculty of Physics and Applied Computer Science - al. Mickiewicza 30, 30-059 Krak\'ow, Poland.}
\address[pan]{Institute of Nuclear Physics, Polish Academy of Sciences - ul. Radzikowskiego 152, 31-342 Krak\'ow, Poland.}
\address[aus1]{GRAPES - rue de la Belle Jardiniere, B-4031 Liege, Federation Wallonie-Bruxelles, Belgium.}
\address[aus2]{e-Humanities Group, KNAW - Joan Muyskenweg 25, 1096 CJ Amsterdam, The Netherlands.}
\address[aus3]{School of Management, University of Leicester - University Road, Leicester, LE1 7RH, UK.}


\begin{abstract}
We report investigations on the statistical characteristics of the baby names given between 1910 and 2010 in the United States of America. For each year, the 100 most frequent names in the USA are sorted out. For these names, the correlations  between the names profiles are calculated for all pairs of states (minus Hawaii and Alaska). The correlations are used to form a weighted network  which is found to vary mildly  in time. In fact, the structure of communities in the network remains quite stable till about 1980. The goal is that the calculated structure approximately reproduces the usually accepted geopolitical regions: the North East, the South, and the "Midwest + West" as the third one. Furthermore, the dataset reveals that the name distribution satisfies the Zipf law, separately for each state and each year, i.e. the name frequency $f\propto r^{-\alpha}$, where $r$ is the name rank. Between 1920 and 1980, the exponent $\alpha$ is the largest one for the set of states classified as 'the South', but the 
smallest one for the set of states classified as  "Midwest + West". Our interpretation 
is that  the pool of  selected names was quite narrow in the Southern states. The data is compared with some related statistics of names in Belgium, a country  also with different regions, but having quite a different scale than the USA. There, the Zipf exponent is low for young people and for the Brussels citizens.
\end{abstract}


\end{frontmatter}

\section{Introduction}

In sociophysics/quantitative sociology, any idea of a research project is inextricably interwoven with an access to related data. The data on babies' names in the United States are available \cite{data} for a long time interval (1880-2011), i.e.  the second half of the stretch of history of the USA since the American independence. Thus, they give a unique opportunity to investigate cultural trends in this large country throughout several  decades  of years. In \cite{wentian}, some related data has been analysed from the perspective of the Zipf law, i.e. the name popularity $y$ vs. the name rank $r$. The Zipf law ($y\propto r^{-\alpha}$) has been found  to be valid only in the  large $r$ rank range. The time dependence of the parameter $\alpha$ has been found to show a weak and wide maximum in the early 50's, slightly more visible for boys' names. In \cite{my}, the data has been analysed in terms of the theory of fashion \cite{simmel}. Two conclusions have been highlighted in  \cite{my}:  (i) for many names, 
the rise of 
popularity is more abrupt than its fall; (ii) the time interval in which a name is  popular is shorter when the set of selected names is richer. The results of the data analysis in \cite{my,chi} indicated that this richness (measured by an index termed 'inequality' in \cite{chi} and 'fragmentation' in \cite{my}) increases in time; yet, the data are not monotonous again ca. 1950. In \cite{chi}, a model has been formulated based on the  social impact for a family, resulting from the choice of the name of the newborn baby. It was concluded that this impact decreases in time. In \cite{pari}, the same American data has been 
investigated in  different states, in 1910-2012. In particular, the Pearson correlations have been calculated for the names frequencies for all states and for each year. The  methods applied in \cite{pari} indicate that the southern and northern states form two uncorrelated clusters, which persist until 1960. Also, the results of a Principal Component Analysis indicated that the difference between the first and second eigenvectors is the largest in 1950; thereafter, the partition is  found to be the sharpest. Finally, the time dependence of the popularity, when averaged over the names, confirmed that the popularity decay  goes more  slowly than its rise (Fig. 6 in \cite{pari}). \\

In the present report, our aim is to use the results of the correlations of the names popularity between different states, to an identify communities in the network of states. These results are combined with the time and state dependent Zipf indices $\alpha$ in the low rank ranges. These tasks are described in the three subsequent sections. Next,  we provide some original information on related data on Belgian (BE) names, - Belgium being a country  also with different regions, as the USA,  but having a quite different smaller population and area size scales than the USA. The BE data, although over a less complete time intervals, allow to validate our conclusions, given in the last section.

\section{Correlations between states}

For each year in the time interval [1910-2011], the set of most popular $N=100$ names  in 48 USA states (we have no data from Alaska and Hawaii) has been  selected  \cite{data}. For each of these states and for each of these names, the percentage $p$ of newborn babies with this given name has been derived.  Then, the contribution of the $a$ state to the popularity of the name $i$ in the year $t$ is found from ,
 
\begin{equation}
x(i,t,a)=\frac{p(i,t,a)}{\sum\limits_{b =1}^K p(i,t,b)}.
\label{def}
\end{equation}
where $K=48$ is the number of states. The  (name) mean of the variable $x$ as a function of time and state  is

\begin{equation}
\langle x(t,a) \rangle =\frac{1}{N}\sum_{i=1}^N x(i,t,a).
\label{mom}
\end{equation}
With the deviation from the mean defined as $y(i,t,a)=x(i,t,a)-\langle x(t,a) \rangle$ and its variance $\sigma ^2(t,a)= \langle y^2(t,a) \rangle$, the Pearson correlation coefficients $\rho$ of correlations between the states $a$ and $b$ are 

\begin{equation}
\rho(t,a,b)=\frac{\sum\limits_{i=1}^N y(i,t,a)y(i,t,b)}{N\sqrt{\sigma^2(t,a)\sigma^2(t,b)}}.
\label{cor}
\end{equation}

The variable $y$ has been so defined as to cancel differences in frequencies of the names from the selected set, a fluctuation which  can be remarkably large as follows from our perusal of the data.\\

Recall that our purpose is to emphasize  whether there are popular reactions in whatever states on particular name trends. The ($a,b$) correlation matrix  represents a weighted network, where the states are nodes and the correlations give the weights of links. This network is analyzed in the next section.

\section{Communities of states}

Since the correlation coefficients are in the range $(-1,1)$, the weighted matrix $w_{a b}(t)$  can be  constructed from  $w_{a b}(t)=(1+\rho(t,a,b))/2$ such that $w \in $ [0,1]. The communities of the network of states are identified from  the set of differential equations \cite{mk1}

\begin{equation}
\frac{dw_{a b}(t)}{dt}=G(w_{a b}(t))\sum_{c=1}^{K-2}(w_{a c}(t)w_{c b}(t)-\beta),
\label{dif}
\end{equation}
where $G(x)=\Theta(x)\Theta(1-x)$, $\Theta (x)$ is the Heaviside step function and $\beta$ is a free parameter. The statistical significance of the obtained partition of the network can be evaluated by calculating the modularity $Q$ \cite{mej},

\begin{equation}
Q= \frac{1}{m}\sum_{ab}(w_{ab}-k_ak_b/m)\delta(c_a,c_b)'
\label{QQ}
\end{equation}
where $k_a$ - weighted degree of the node (state) $a$, and $m=\sum_{ab}w_{ab}$. The value of the parameter $\beta$ is chosen by trials as to obtain the maximal modularity $Q$. The latter condition serves also as a criterion to point on which partition is the most significant one. The method has been validated in \cite{mk1,mk2}.\\
 
\begin{figure}[!hptb]
\begin{center}
\includegraphics[width=\columnwidth]{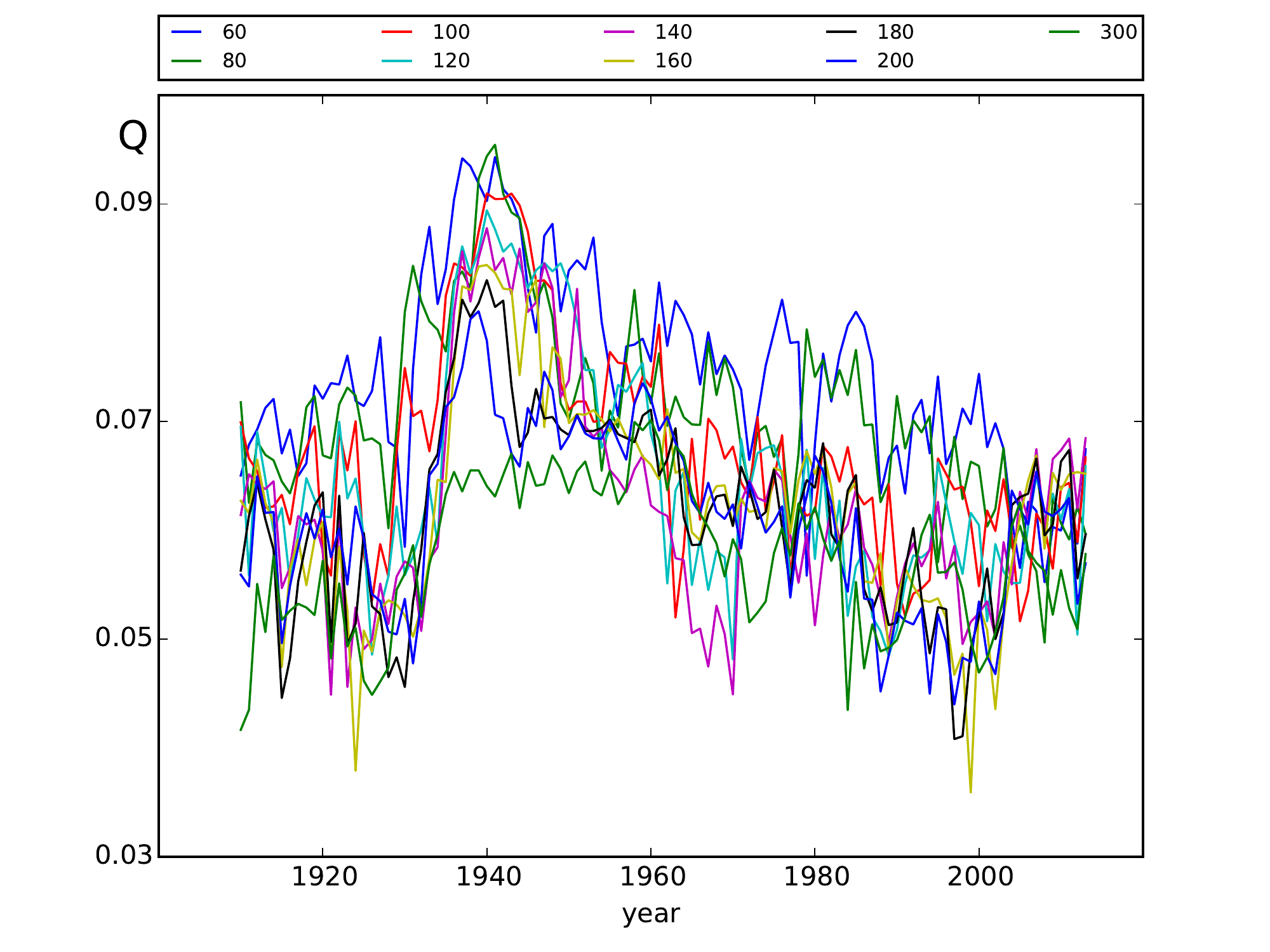}
\caption{The modularity $Q$ of the obtained partitions vs. time, for different numbers $N$ of names. The largest values of $Q$ are obtained for $N$ between 60 and 100. However, for $N=60$ the fluctuations of $Q$ are larger.}
\label{Q}
\end{center}
\end{figure} 
 
\begin{figure*}[!hptb]
\begin{center}
\includegraphics[width=\textwidth]{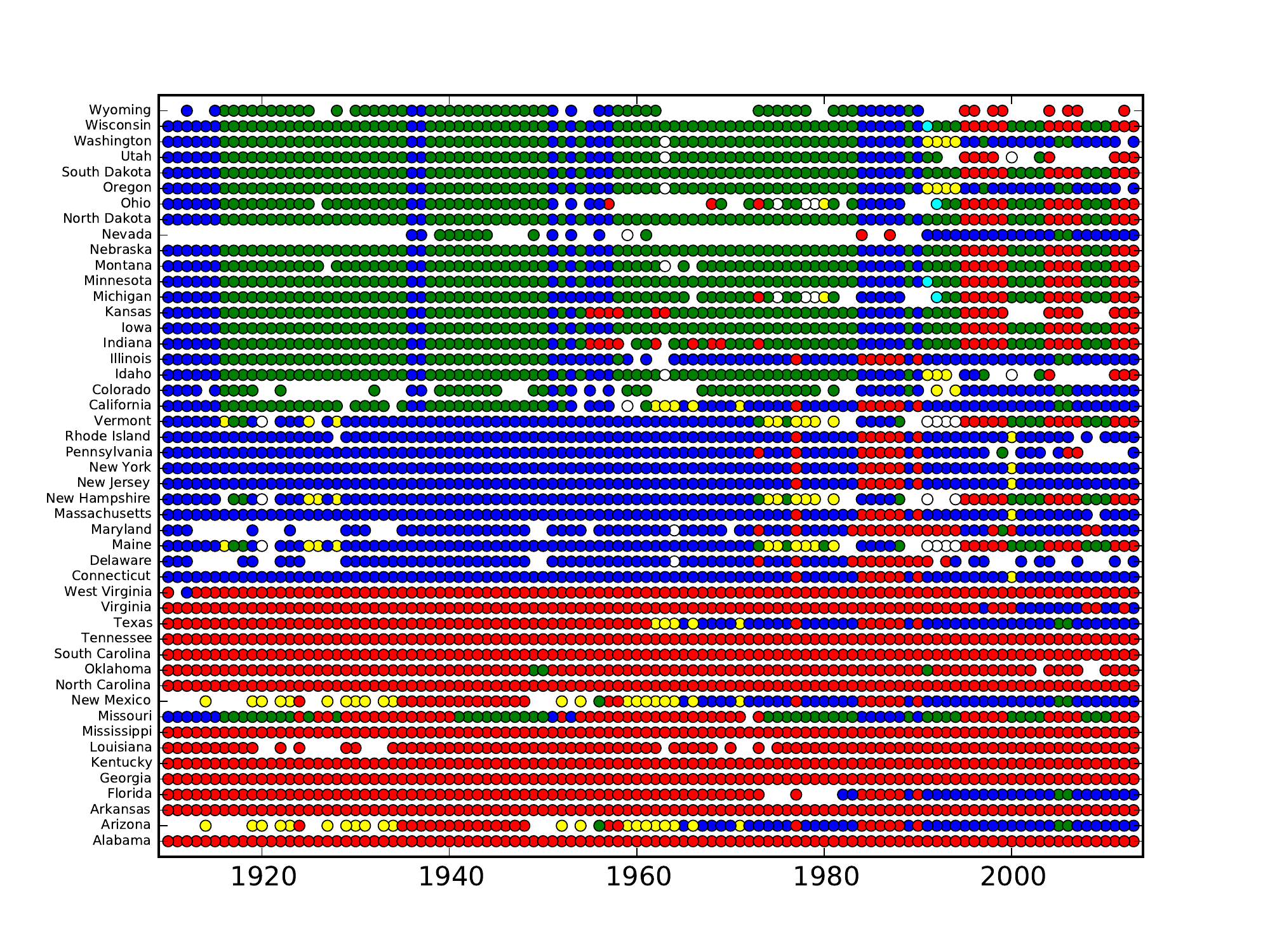}
\caption{The communities of the states vs. time. States which belong to the same communities are marked with the same color (color online).}
\label{comm1940}
\end{center}
\end{figure*}

For the obtained partition of states, the modularity remains very low, about 0.06; it is only ca. 1940 that a clear maximum of $Q$ exceeds 0.09 (see Fig. (\ref{Q}). Yet, the obtained partition is surprisingly stable, as shown in and deduced from Fig. (\ref{comm1940}).  In this illustration, the states which are assigned to the same "community" are marked with the same color. The colors are so chosen as to obtain the maximal overlap with the colors for 1940, when the modularity is maximal. \\

As seen in Fig. (\ref{comm1940}), the community structure becomes more fuzzy after ca. 1980. Yet three main corpora of states can be distinguished. The content of these groups is quite consistent with the commonly admitted geo-political regions of  the USA,  taken after \cite{reg}. To be specific: \\
- all 9 states of the NorthEast are properly assigned to one community;\\
-  18 states of the Midwest and the West are properly assigned to asecond community. However,the status of Arizona, California, Missouri, New Mexico and Nevada varied in time, so their assignment remains unclear; \\
- 13 states of the South are properly assigned to a  third community. However, the status of Texas remains unclear for the same reason as above. Also, Delaware and Maryland are assigned to the NorthEast.\\

 The above classification of the states to communities, based on the data on names, is not far from being distinguished from  the "content "of the administrative regions. The largest departure is related to five states in the Southern West (California, Nevada, Arizona, New Mexico and Texas), which admittedly can be considered to share some common cultural characteristics. In three other cases (Delaware, Maryland and Missouri), the proximity with the other neighboring region also allows for considering some cultural influence.\\

This partition can be verified by a comparison with the results obtained by means of the spectral algorithm by Newman \cite{speal}. These results are quite similar. To list the differences: the results obtained with the spectral algorithm remain stable also after 1980. Further, Maryland is assigned to Northeast, and California and Nevada belong to Western states more firmly. On the other hand, Kansas and Indiana appear to join the South between late 50's and early 70's. Also, Arizona is attached to Northeast from 1960 till 2010. We should add that a strict accordance of the obtained partition with administrative regions is not an ultimate criterion of its quality.\\

 \begin{figure*}[!hptb]
\begin{center}
\includegraphics[width=\textwidth]{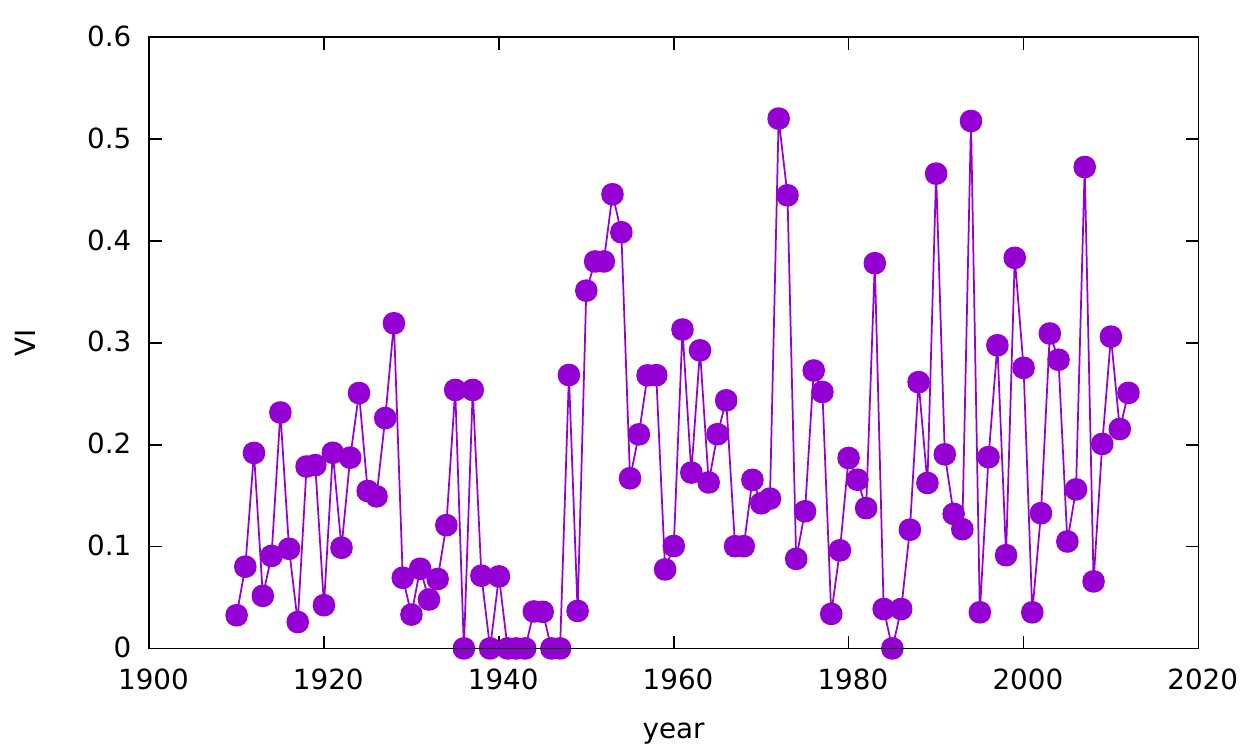}
\caption{The index of variation of information (VI) vs. time.}
\label{varr}
\end{center}
\end{figure*}

The time dependence of the structure of communities can be traced by a comparison of this structure in each year with the one in the next year. The index of this variation (variation of information, VI) \cite{mei} is a measure of information gained or lost when one partition is transformed into another. In Fig. (\ref{varr}) we show the time dependence of VI. It is interesting that between 1939 and 1947, this index is close to zero. This fact coincides with the maximum of the modularity $Q$, shown in Fig. (\ref{Q}). In other words, the obtained partition of states into communities is simultaneously most clear and most stable around 1940.

\section{The Zipf exponent}

 \begin{figure}[!hptb]
\begin{center}
\includegraphics[width=\columnwidth]{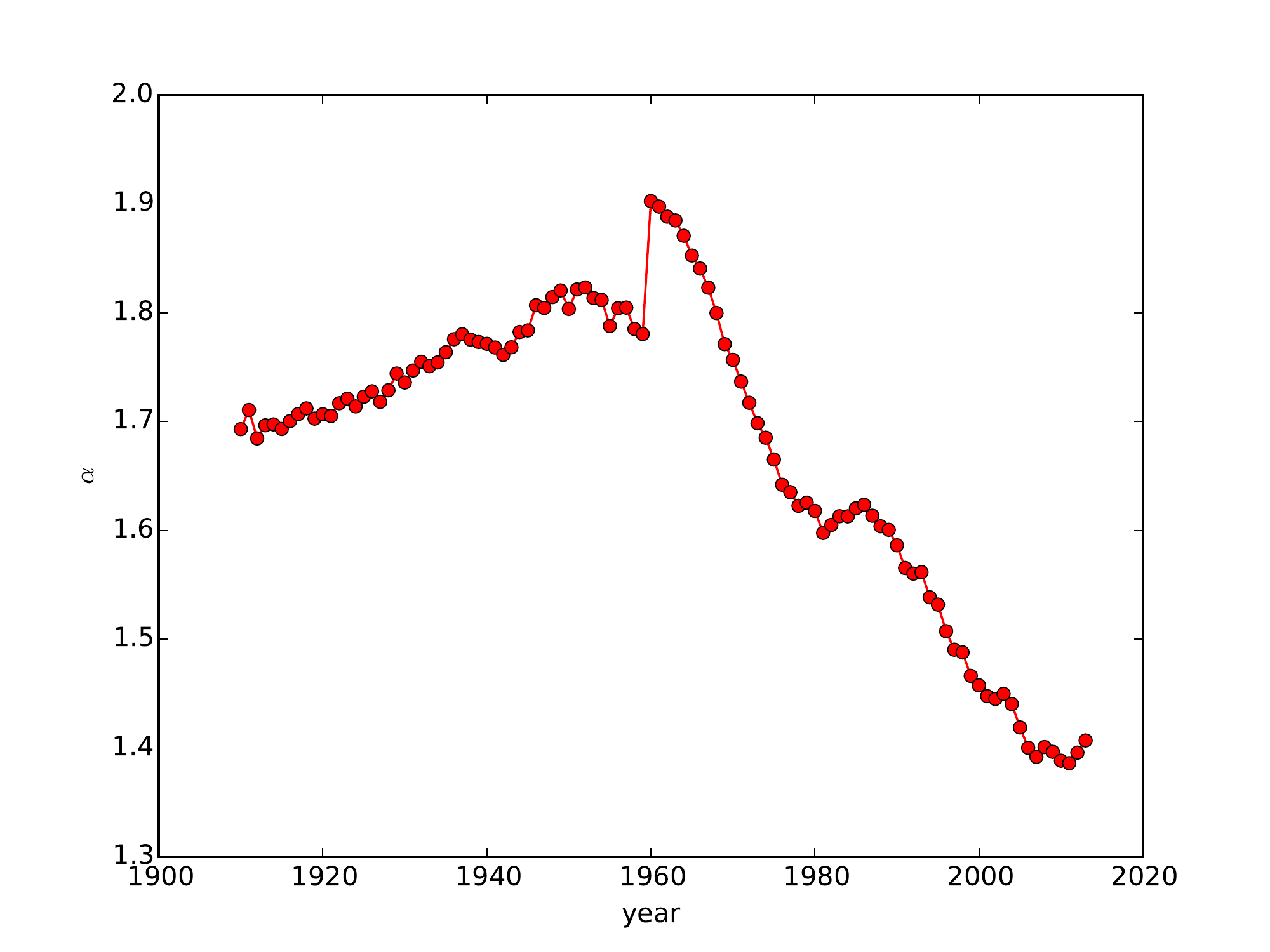}
\caption{The Zipf exponent $\alpha$, averaged over all states, vs. time.}
\label{atot}
\end{center}
\end{figure}

\begin{figure}[!hptb]
\begin{center}
\includegraphics[width=\columnwidth]{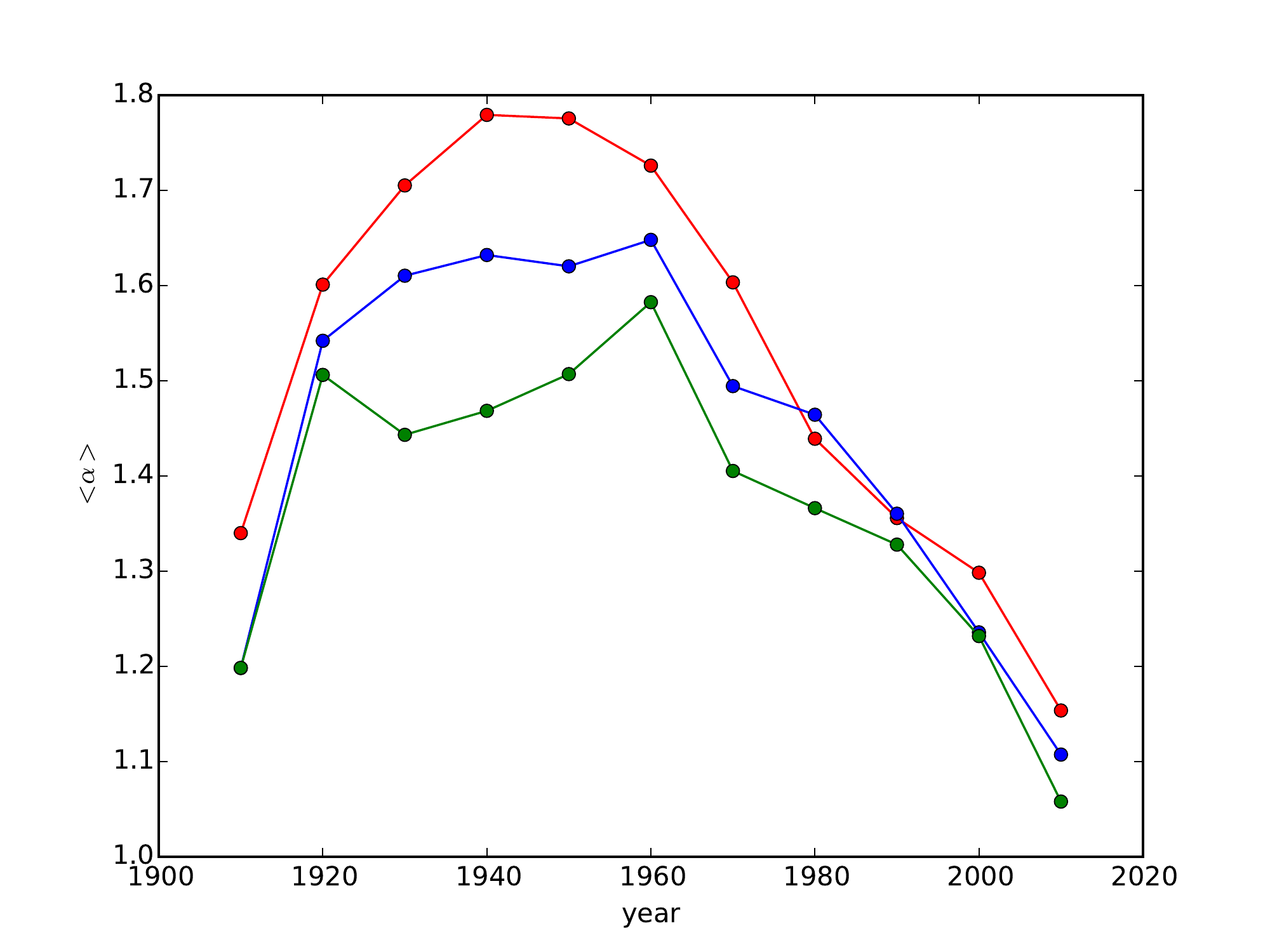}
\caption{The Zipf exponent $\alpha$, averaged over the states within each of three main communities, vs. time. Between 1920 and 1980, the largest value of $\alpha$ is for the Southern states (red line, color online), and the smallest - for the Western states (green line).}
\label{a3}
\end{center}
\end{figure}

Since we analyze only the most popular names,  to compare directly the sizes of the full sets of baby names in different years or in different states is out of the paper purpose. However, the Zipf rank-frequency plots  can be used to assess how rich  those sets are  when comparing the corresponding Zipf exponents $\alpha$ (if such exponents can be estimated). Similar to the interpretation of the word-frequency plots in linguistics, the smaller is $\alpha$, the larger is the diversity of analyzed items~\cite{ferrer,ggkkd}. In the  present case,  the time dependence of the Zipf exponent $\alpha$, when averaged over all states, as shown in (Fig.\ref{atot}).  clearly  presents  a  maximum ca. 1960.  In Fig. (\ref{a3}),  the same data is shown, but separately for the three communities of states identified in the previous section. 

Three features are to be highlighted here:  the first is a relevant remark usual on nonlinear systems;  the next two other points  though  only qualitative statements are nevertheless worth of consideration.  First, the data shown in Fig.(\ref{atot}) cannot be treated as an average of the data shown in Fig. (\ref{a3}). Second, 
the sharp maximum of $\alpha$ seen in 1960 in the national data (Fig. \ref{atot}) is related to two communities: the West and the Northeast, but not to the South. In the latter, there is a broader maximum between 1940 and 1950. Third, the order of the values of the Zipf exponents between 1920 and 1980 obeys the rule: $\alpha$(the West) $<$ $\alpha$(the Northeast) $<$ $\alpha$(the South). This means, that the values of $\alpha$ calculated for particular states of the Northeast are systematically larger than the values for particular states of the West; on the other hand, such $\alpha$ values are systematically smaller than those for particular Southern states.\\

\section{The Belgium data}

Having found some network structure and systematic organization of names between states and regions in the USA, it is of interest to consider some reference data for a comparison.  Therefore, in this section,  we discuss the Zipf law for additional data about  names \cite{MA} found in Belgium. The data are less complete  in time than for the USA:  however,  they allow to evaluate the Zipf index  distinguishing for gender, i.e. for the male and female names in Belgium, furthermore also separately for the Flemish Region (Flanders), the Walloon Region (Wallonia) and the Brussels-Capital Region (Brussels).  Thus the data has some similarity with  that used  for the USA investigation. However, the data is not related to newborn babies, but to the cohorts of age in given ranges: {\it (i)} less than 18, {\it (ii)} between 18 and 64, {\it (iii)} over 65, leading to some other complementary perspective as well. 
The related Zipf exponents are given in Table  \ref{BEmfZipf}.


\begin{table}[htp]
\caption{The values of the Zipf exponents $\alpha$ for Belgium, according to the regions and age range. Asterisks mark the results, where fitting was possible only with the last part of the tail, - one order of magnitude of the rank} 
\label{BEmfZipf}
\begin{center}
\begin{tabular}{|c|c|c|c||c|c|c|c|c|c|c|c|}
\hline
&males&females\\
\hline
Belgium&1.62&1.67\\
Flanders&1.67&1.57\\
Wallonia&1.60&1.69\\
Brussels&1.36&1.37\\\hline
Belgium $<18$&1.42&1.54\\
Belgium $18-64$ &1.65&1.72\\
Belgium $>65$ & 1.85&1.83\\ \hline
Flanders $<18$&1.41&1.33\\
Flanders $18-64$ &1.65&1.60\\
Flanders $>65$ & 2.18*&2.03\\ \hline
Wallonia $<18$&1.53&1.42\\
Wallonia $18-64$ &1.56&1.64\\
Wallonia $>65$ & 1.58&1.66\\ \hline
Brussels $<18$&1.41*&1.30*\\
Brussels $18-64$ &1.37&1.33\\
Brussels $>65$ & 1.32&1.40\\
 \hline			
\end{tabular}
\end{center} \end{table}

At this point, some semi-qualitative conclusions seem justified. These are as follows. No systematic difference is distinguished between both sexes. Neither  do we see meaningful differences between Wallonia and Flanders. The data for the youth in the whole Belgium are slightly smaller than for the two other groups also in the whole Belgium, but when applied to the regions, this rule does not hold. For the Brussels, the $\alpha$  values are systematically smaller than for the other regions, indicating a tendency toward a more uniform distribution. This might be due to the heterogeneity of the population.

\section{Discussion}

Nowadays, many topics are based on network models,  spanning biological, ecological, and economic systems  as well as cultural systems \cite{PRL86.01.5835_wghtevnetwk,PRE74.06.26102LangNetw_rodgers,gligorausloosEPJB,gligorausloosAPPA,PNAS107.10.13636-41-Szell-RLambi-SThur,QuQu48.14.1893ITnetwkGRamda,QuQu48.14.leadernetwkmerlone}. Most of these are best described by weighted networks.  Moreover, physical models attempt to take into account the rationality (or/and irrationality) behavior of agents in community structures \cite{BJP42AtmanGoncalves}. Both aspects have been taken into account in the present investigation. 

It is also  logical to admit that  an agent behavior is often  influenced by the action of others, due or not to observation and modeling. The Zipf method of data analysis is likely one, if not the quickest, way of introducing modeling considerations after data analysis. Thus,
it is tempting to state that the systematic differences found between the values of the Zipf exponent for baby names are due to cultural differences between given regions. A large value of $\alpha$ may originate from a shorter set of socially accepted names (which suggests stronger social pressure), strict patterns of naming children within families, or stronger cultural isolation. Actually, this is the basis of the model, proposed in \cite{chi} and thereby confirmed. The results on the American baby names indicate that this kind of pressure was relatively strong in the Southern region, where the rural tradition has been described broadly as the Black Belt culture \cite{joh}. On the other hand, the Belgium data suggests that small values of the Zipf exponent  are  correlated with  strong urbanization. However, this conclusion is not entirely consistent with the results shown in Fig. (\ref{a3}) for the USA, in which the  West show lower values of $\alpha$ than New York and the states of New England. One way 
or 
resolving the anomaly maybe in considering some more heterogeneously distributed culture. The issue, why any kind of social pressure could be  smaller in the Western states than in the NorthEast, nevertheless remains somewhat open.\\

Some hint can be provided by the migration data. As thoroughly described in \cite{jng}, the internal migration in the USA, was  intensive in the first decades of the 20th century, was temporarily slowed  down in  the 30's because of the economic crisis, but  largely accelerated again in the 40's with the outbreak of WWII. Accordingly, the data on the intensity of internal migration shows a jump in the 40's \cite{mol}. Accepting that the richness of the datasets of  baby names increases because of these migrations, this could throw some light also on the change in trend of the fragmentation index of the baby names (termed as inequality in \cite{chi}), observed shortly after 1940 \cite{my,chi}. On the other hand, the time dependence of the modularity $Q$ shows also a maximum near 1940. Recalling that $Q$ is a measure of a statistical significance of the obtained partition of the network into communities, we are tempted to suppose that cultural boundaries between the regions were particularly 
rigid in 1940, just before the acceleration of the migration processes \cite{jng} and an ethnic/cultural spreading. \\

After 1980, the migration processes are weakening \cite{jng}. However, the decrease of the Zipf exponent continues, accompanied by a similar decrease of the fragmentation index \cite{my,chi}, more or less in the same way in all three communities of states. At this time, the process of cultural mixing in different states seems saturated. On the other hand, the liberalization in the selection of baby  names continues, plausibly driven by media and  significant immigration from abroad. The latter explanation is supported, for example, by comparing the sets of the most popular names in the 60's  with more recent ones, when the appearance and increasing  number of the names of Hispanic origin can be observed in the top 100 list  for many states \cite{data}.\\

If these conclusions are accepted, the data on Belgium  names appear to be coherent with the hypothesis on the liberalization. The richest the dataset, the smallest Zipf exponent $-$ this rule is consistent with the fact that the Zipf exponent is small both for the youngsters and for the Brussels region. \\

Summarizing, the correlations between frequencies of given names in different states are used to form a weighted network. The communities in this network, although time-dependent, remain surprisingly stable till ca. 1980, and their structure matches the administrative and cultural regions of the USA. Also, the time- and state-dependent Zipf exponents, when averaged over the obtained communities of states, allow to infer about the cultural pressure in the related regions.

\vspace{1cm}
\noindent
{\bf Acknowledgement}\\
We are grateful to Dorota Prasza{\l}owicz and to our Anonymous Referee for helpful comments. This paper is part of scientific activities in COST Action TD1210 'Analyzing the dynamics of information and knowledge landscapes' and COST Action IC120 'Computational Social Choice'. The work was partially supported by the Polish Ministry of Science and Higher Education and its grants for Scientific Research and by the PL-Grid Infrastructure.


\end{document}